\documentclass[%
 reprint,
 nobalancelastpage,
 superscriptaddress,
 amsmath,amssymb,
 aps,
 prb,
]{revtex4-2}

\usepackage{times,comment}
\usepackage{multirow}
\usepackage{newtxmath}
\usepackage{easybmat}% dashed line for block matrix
\usepackage{graphicx}% Include figure files
\usepackage{dcolumn}% Align table columns on decimal point
\usepackage{bm}% bold math
\usepackage{hyperref}% add hypertext capabilities
\usepackage{xcolor}
\usepackage{verbatim}

\begin{document}

\preprint{APS/123-QED}

%TC:ignore
%\title{Non-Hermitian Dirac Vortex: Application to Topological-Cavity Surface-Emitting Laser}

\title{Non-Hermitian Dirac Vortex: Minimal Theory for Topological-Cavity Surface-Emitting Laser}
%\title{Non-Hermitian Dirac Vortex: analytical theory for Topological-Cavity Surface-Emitting Laser}

\author{Zong-Liang Li}
\affiliation{Institute of Physics, Chinese Academy of Sciences/Beijing National Laboratory for Condensed Matter Physics, Beijing 100190, China}
\affiliation{School of Physical Sciences, University of Chinese Academy of Sciences, Beijing 100049, China}
\author{Guang-Rui Li}
\affiliation{Institute of Physics, Chinese Academy of Sciences/Beijing National Laboratory for Condensed Matter Physics, Beijing 100190, China}
\author{Le-Chen Yang}
\affiliation{Institute of Physics, Chinese Academy of Sciences/Beijing National Laboratory for Condensed Matter Physics, Beijing 100190, China}
\author{Zhong Wang}
\affiliation{Institute for Advanced Study, Tsinghua University, Beijing, 100084, China}
\author{Ling Lu}
\email{linglu@iphy.ac.cn}
\affiliation{Institute of Physics, Chinese Academy of Sciences/Beijing National Laboratory for Condensed Matter Physics, Beijing 100190, China}

%\date{\today}

\begin{abstract}
We construct a non-Hermitian Dirac-vortex model that combines a complex-mass winding with an infinite-imaginary-potential boundary, extending the Jackiw-Rossi and neutrino-billiard models to the dissipative regime. Moreover, this model serves as a minimal theory for the recently proposed topological-cavity surface-emitting laser~(TCSEL): the imaginary mass encodes vertical radiation loss and the absorbing boundary defines the active region. We derive closed-form expressions for the modal frequencies, thresholds, and tunable vector-beam polarizations, which are validated experimentally. Our work provides a rare example in which an analytical non-Hermitian topological theory captures the essential physics for engineering practical optoelectronic devices.

%Suggested reviewers: Thomas, Berry, Inoue, Akzyanov, Guancong Ma, Li Ge, Chamon, Huang Liang, Peng Chao, Xuefan Yin，Dezhuan Han.

\end{abstract}

%TC:endignore

\maketitle

\textit{Introduction --- }
Relativistic Dirac particles cannot be confined by real potentials, even of infinite height, because of Klein tunneling. Confinement has nevertheless been achieved through spatial mass profiles. In the MIT bag model~\cite{Chodos1974}, an infinite Dirac-mass well confines quarks, a mechanism later adapted to two dimensions~(2D) in studies of neutrino billiards~\cite{berry1987neutrino} and graphene disks~\cite{Akhmerov2008,Christensen2014}. A distinct class of confinement arises from topological defects in the mass field---such as kinks~\cite{Jackiw1976}, vortices~\cite{Jackiw1981}, or monopoles~\cite{Cheng2024}---which underlie Majorana bound states~\cite{Fu2008,Teo2010} in condensed-matter systems. Here, we extend these seminal models into the non-Hermitian regime without losing their analytical solvability. Specifically, we solve for finite-lifetime bound states in an infinite well of imaginary potential with vortex configurations of both real and imaginary Dirac masses.

Our non-Hermitian Dirac-vortex model provides a minimal theory for the emerging topological-cavity surface-emitting laser~(TCSEL)~\cite{gao2020dirac, yang2022topological, ma2023room, han2023photonic, Liu2024, Zhong2026}, whose stable single-mode lasing is enabled by the vortex zero mode. TCSEL generalizes the 1D kink-mode lasers~\cite{morthier2013handbook, padullaparthi2021vcsel} used in everyday technologies such as cell phones and internet communications, while extending the 2D photonic-crystal surface-emitting lasers~(PCSELs)~\cite{yoshida2023high,Noda2024} from periodic structures to lattices with topological defects. The modeling of these large-area devices, spanning thousands of lattice periods in diameter, is computationally prohibitive for full-wave simulations~\cite{gao2020dirac, ma2023room,Liu2024,Zhong2026, han2023photonic, gao2025far} and therefore relies on the numerical coupled-wave theory~(CWT)~\cite{Kogelnik1972, Kazarinov1985, Liang2012, Liang2013three, Inoue2022general} that has been continuously developed over the past 50 years. In this letter, we derive analytical solutions to the non-Hermitian Dirac Hamiltonian of TCSEL, including the frequencies, losses, and radiation patterns of the modes, that reveal the mechanisms of single-mode operation in finite topological cavities.

%%%%%%%%%%%%%%%%%%%%%%%%%%% Section_2 %%%%%%%%%%%%%%%%%%%%%%%%%%%

\textit{Complex-mass vortex--- }
We extend the two mass terms, forming the Dirac vortex in the Jackiw-Rossi model, from real to complex values. 
The resulting non-Hermitian $4\times4$ bulk Hamiltonian and its eigenvalues are given by Eq.~\eqref{eq:dirac_H}, which can be derived from the coupled-wave theory as detailed in Supplemental Material~[43], Sec.~I.
\begin{equation} 
\label{eq:dirac_H}
\begin{split} 
H_0(\boldsymbol{k}) &= -\frac{1}{2} \tau_z \boldsymbol{\sigma}\cdot\boldsymbol{k}  + (m_1 + i \mu_1) \tau_x + (m_2 + i \mu_2) \tau_y + i \mu \\
\omega + i\frac{\alpha}{2} &= i\mu \pm \sqrt{(m_1+i \mu_1)^2+(m_2+i \mu_2)^2+(\frac{k}{2})^2},
\end{split}
\end{equation}
where the real-valued $m_{1,2}$ and $\mu_{1,2}$ are the real and imaginary mass terms, and $\mu=\sqrt{\mu_1^2+\mu_2^2}$ ensures that the system remains passive without gain under the $e^{i\omega t}$ time-harmonic convention.
$\boldsymbol{k}=(k_x,k_y)$ is the momentum vector, and $\sigma_i, \tau_i$ are the Pauli matrices. 
The factor of $1/2$ in the kinetic term is the group velocity of the Dirac cone, while the factor of $1/2$ in the complex eigenvalue ensures that $\alpha$ is the decay rate of the field intensity~(rather than amplitude).
%, with the time-harmonic convention $e^{i\omega t}$.

\begin{figure}
\centering
\includegraphics[width=0.98\linewidth, keepaspectratio]{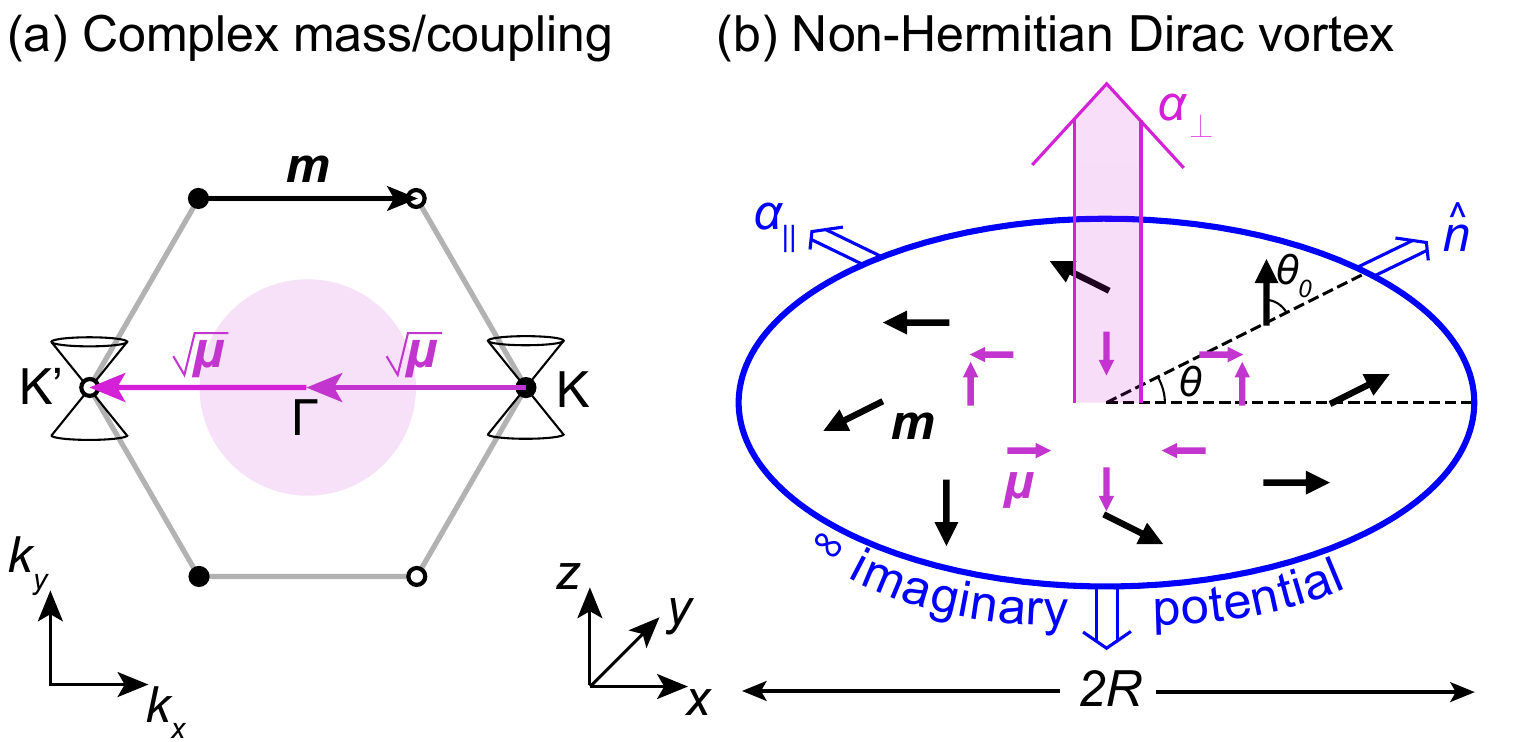}
\caption{Non-Hermitian Dirac vortex. (a) Momentum-space couplings between K (solid dots) and K$'$ (hollow dots) valleys in the Brillouin zone of a triangular lattice. The central pink circle represents the radiative light cone. The black arrow indicates the direct Hermitian coupling $\boldsymbol{m}$, while the pink arrows labeled $\sqrt{\boldsymbol{\mu}}$ denote the indirect non-Hermitian coupling via the vertical radiation channel at the $\Gamma$ point. (b) Real-space view of the cavity with radius $R$. The Hermitian mass vector $\boldsymbol{m}$ (black arrows) forms a vortex with winding number $+1$, whereas the non-Hermitian mass vector $\boldsymbol{\mu}$ (pink arrows) forms a vortex with winding number $-2$ (due to the double imaginary couplings of opposite direction with the real coupling). The system is confined by an infinite imaginary potential (blue color) leading to in-plane boundary loss $\alpha_\parallel$, while the imaginary mass induces vertical radiation loss $\alpha_\perp$.}
\label{fig:coupling}
\end{figure}

Both the real and imaginary masses arise from supercell-induced couplings between the Dirac cones in a $C_3$-symmetric triangular photonic-crystal slab. 
The real mass $m$ arises from the first-order direct in-plane coupling between the cones at the Brillouin-zone boundary~(K, K')~\cite{Hou2007}, indicated by the black arrow in Fig.~\ref{fig:coupling}(a). 
%of the triangular lattice with $C_3$ symmetry~\cite{Hou2007}. 
The imaginary mass $\mu$ arises from the second-order indirect couplings between the cones through the zone center~($\Gamma$), denoted by the two pink arrows in Fig.~\ref{fig:coupling}(a).
$\mu$ is non-Hermitian because $\Gamma$ is inside the light cone, where the states radiate out of plane due to the lack of total internal reflection.
Consequently, $\mu \propto m^2$ and $\mu\ll m=\sqrt{m_1^2+m_2^2}$ due to their different coupling orders.
Moreover, $\boldsymbol{\mu}$ carries twice the coupling phase of $\boldsymbol{m}$ but with opposite signs, due to their opposite coupling directions shown in Fig.~\ref{fig:coupling}(a).

Thus, the Dirac vortex of real masses is accompanied by the spatial winding of imaginary masses.
The mass profiles in polar coordinates~($r$, $\theta$) are
$\boldsymbol{m}(\theta)= m_1+i m_2 = m e^{i(w\theta+\theta_0)}$ and $\boldsymbol{\mu}(\theta) = \mu_1 +i \mu_2 = -\mu e^{- 2i (w\theta+\theta_0)}$,
where $w$ is the winding number of $\boldsymbol{m}$, $\theta$ is the polar angle, and $\theta_0$ is the initial phase of the mass vector $\boldsymbol{m}$ as illustrated in Fig.~\ref{fig:coupling}(b). 
The winding-number difference between $\boldsymbol{m}$ and $\boldsymbol{\mu}$ is $3w$, consistent with the discrete $C_3$ symmetry of the underlying lattice.
To define a finite-size object, we need a proper boundary condition for this complex-mass vortex.

\begin{figure*}[th!]
\centering
\includegraphics[width=1\linewidth]{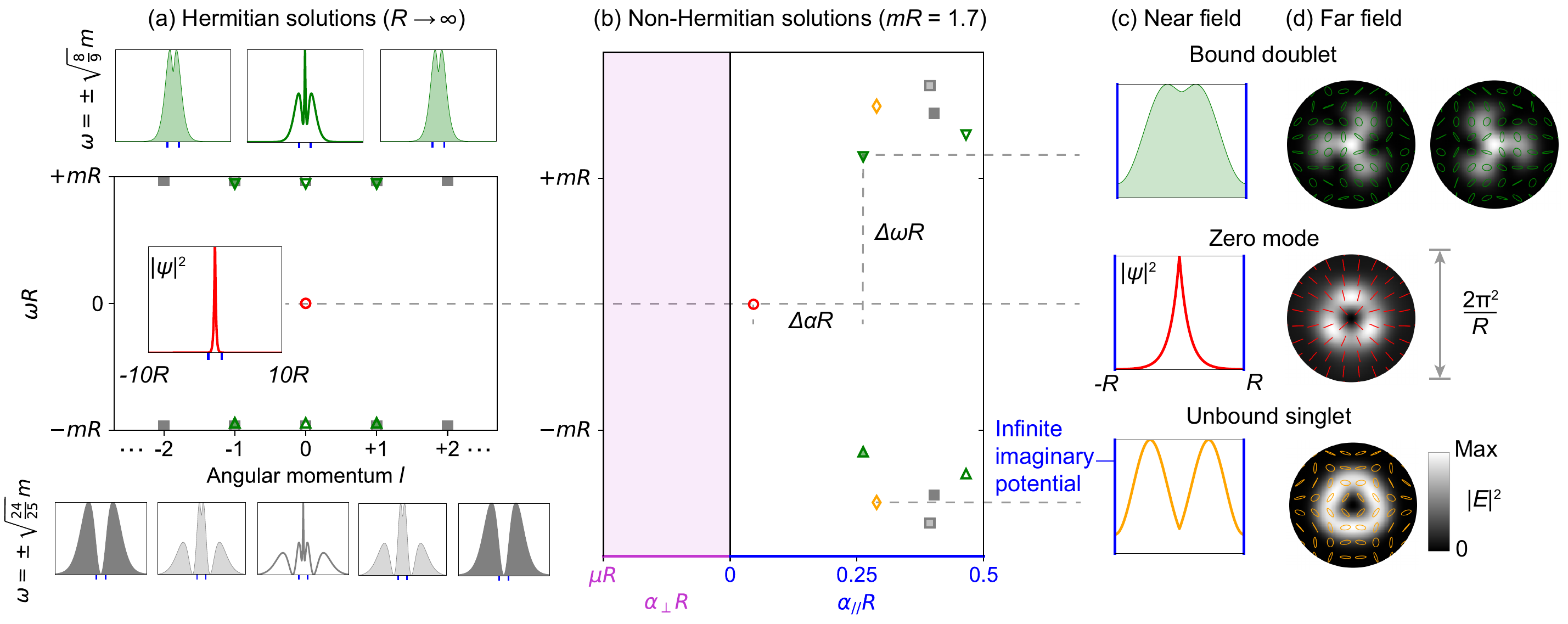}
\caption{\label{fig:spectrum}Eigen-modes to the Hermitian and non-Hermitian vortices. (a) Real spectrum of the Hermitian vortex, showing an infinite series of bound states indexed by principal quantum number $n$ and angular momentum $l$. (b) Complex spectrum of a finite-sized vortex ($mR=1.7$). The horizontal axis is the total loss, consisting of the identical vertical radiations $\alpha_\perp$ (left) and distinct boundary loss $\alpha_\parallel$ (right) among the modes. The zero mode has the minimum $\alpha_\parallel$, securing the single-mode stability margins $\Delta\alpha R$ and $\Delta\omega R$. (c) Near-field radial intensities. The topological mode is localized away from the imaginary potential at the boundary (blue lines), whereas the other modes extend into the absorbing boundary. (d) Far-field vector-beam patterns with polarization directions indicated by colored lines.}
\end{figure*}

\textit{Infinite imaginary potential --- }
We assign the absorbing boundary condition by imposing an infinite imaginary potential outside the vortex~($r>R$).
Changing the infinite potential from real to imaginary does not affect the boundary reflectivity, but it avoids Klein tunneling because a wave decays inside an imaginary potential.
The decay length vanishes when the height of the imaginary barrier diverges, allowing a compact boundary condition that, as shown in Supplemental Material~[43], Sec.~I, can be formulated as Eq.~\eqref{eq:boundary}. 
\begin{equation}
\label{eq:boundary}
\begin{gathered}
H\left(m e^{i (w\theta+\theta_0)}, -\mu e^{-2i( w\theta+\theta_0)}\right)|\psi(\boldsymbol{r})\rangle =(\omega+i\frac{\alpha}{2})|\psi(\boldsymbol{r})\rangle \\[5pt] 
\text{at } r=R,\quad
-\tau_z(\boldsymbol{\sigma} \cdot \hat{n}) |\psi(\boldsymbol{r})\rangle=|\psi(\boldsymbol{r})\rangle
\end{gathered}
\end{equation}
Inheriting the reflection characteristics of Klein tunneling~\cite{Katsnelson2006}, this boundary exhibits minimal reflectivity at normal incidence~(which vanishes for massless particles) and total reflection at grazing angles.
Similar boundaries have been discussed in perfectly matched layers~\cite{Antoine2017} and in the detection theory~\cite{Tumulka2016} of Dirac systems, resembling the scattering boundaries in electromagnetics~\cite{Engquist1977}.

This absorbing boundary condition is appropriate for TCSEL, where the unpumped region outside the cavity has a large absorption coefficient. For quantum wells, the absorption loss is \(\sim 600~\mathrm{cm}^{-1}\) at \(\sim 1~\mu\mathrm{m}\) wavelength~\cite{Coldren2012, Inoue2022general}, much larger than the momentum and mass scales in Eq.~\eqref{eq:dirac_H}, both of which scale as \(1/R\). For a typical device radius \(R=500~\mu\mathrm{m}\), \(1/R=20~\mathrm{cm}^{-1}\). This disparity~($20~\text{cm}^{-1}\ll600~\text{cm}^{-1}$) justifies the use of an infinite imaginary potential as an effective boundary condition, which substantially simplifies the theoretical treatment.

\textit{Symmetries of non-Hermitian vortex --- }
Spectrally, an odd vortex~(odd $w$) possesses a particle-hole symmetry ~\cite{Kawabata2019} that pairs the eigenvalues into $\pm\omega+i\frac{\alpha}{2}$, leaving only the zero mode unpaired for single-mode lasing.
This pairing symmetry is the anti-$\mathcal{PT}$ symmetry~\cite{Ge2013}, because both the masses and the boundary are anti-symmetric under $\mathcal{PT}$, given that the boundary shape is inversion symmetric.
\begin{equation}
\label{eq:anticommutation}
\{\mathcal{PT}, H(\boldsymbol{r})\} = 0, \quad \mathcal{PT}=\left.\sigma_x \tau_x K\right|_{\theta \to \theta+\pi}
\end{equation}
where $\mathcal{P}=|_{\theta \to \theta+\pi}$ denotes parity symmetry, $\mathcal{T}=\sigma_x\tau_xK$ is the time-reversal operator~($\mathcal{T}^2=+1$), and $K$ is complex conjugation.
When the real mass~($w$) winds an odd number of times, satisfying $\boldsymbol{m}(\theta)=-\boldsymbol{m}(\theta+\pi)$, the imaginary mass winds an even number of times~($-2w$), satisfying $\boldsymbol{\mu}(\theta)=\boldsymbol{\mu}(\theta+\pi)$.
So $m$ is anti-symmetric under $\mathcal{P}$ but symmetric under $\mathcal{T}$, while $\mu$ and the imaginary potential are symmetric under $\mathcal{P}$ but anti-symmetric under $\mathcal{T}$~(loss-gain flip).
Being the only unpaired state at $\omega=0$, the zero mode maps to itself under $\mathcal{PT}$.
Consequently, as a $\mathcal{PT}$ eigenstate, the zero mode emits linearly polarized far-field radiation, while the other paired modes emit elliptically polarized radiation, as shown in Fig.~\ref{fig:spectrum}(d).

Spatially, a mass vortex with uniform angular winding, together with a circular boundary, possesses rotational symmetry that permits analytical solutions via the separation of variables in polar coordinates:
\begin{equation}
\label{eq:separation}
\psi(r,\theta)=e^{i l \theta-i \frac{\sigma_z}{2} \theta} e^{-i \frac{\tau_z}{2} (w\theta+\theta_0)}\boldsymbol{f}(r)
\end{equation}
where $l$ is the angular-momentum quantum number and $\boldsymbol{f}(r)=(f_1,f_2,f_3,-f_4)^{\mathrm{T}}$ is a four-component radial spinor.
The corresponding angular momentum operator is $L_z = (-i\partial_{\theta} + \frac{1}{2}\sigma_z) + \frac{w}{2}\tau_z$, where the first two terms are from the conventional Dirac Hamiltonian while the last term accounts for the vortex texture, so that $[L_z,H(\mathbf{r})]=0$.
However, the real and imaginary masses wind differently and cannot share a common angular momentum operator.
%(In the lattice realization, the continuous rotational symmetry reduces to $C_3$, which is compatible with both winding numbers of +$w$ and -2$w$.)
To obtain analytical solutions in the form of Eq.~\eqref{eq:separation}, we neglect the imaginary mass $\mu$~($\ll m$), which has winding number $-2w$. 
% This approximation is justified as $\mu$ is typically a few $\mathrm{cm}^{-1}$, which is much smaller than the real mass $m$ (typically tens to hundreds of $\mathrm{cm}^{-1}$). 
The radiative properties associated with $\mu$, such as the loss and far field, are calculated perturbatively.
%from the wavefunctions.

\textit{Identical radiation loss --- }
All modes of the non-Hermitian vortex share the same radiative loss~($\alpha_{\perp}$), as shown by first-order perturbation theory.
\begin{equation}
\label{eq:rad_loss}
\alpha_{\perp} = 2 \langle \psi | \mu_1(\theta)\tau_x+\mu_2(\theta)\tau_y+\mu | \psi \rangle = 2\mu.
\end{equation}
Both imaginary coupling terms $\mu_{1,2}(\theta)$ vanish upon angular integration due to the mismatch in winding number between $\boldsymbol{\mu}$~($-2w$) and $\boldsymbol{m}$~($w$), the latter determining the wavefunctions.
Since lasing occurs first in the mode with the lowest total loss~($\alpha_\perp+\alpha_\parallel$), the mode discrimination of TCSEL relies on the in-plane boundary loss~($\alpha_\parallel$), which we solve next.

\textit{Solution of Hermitian vortex --- }
We first review the solutions of the classical Jackiw-Rossi model --- the Hermitian Dirac vortex~($w=1, \mu=0, R\to \infty$).
Since the radial Dirac equation can be mapped to the Schr\"{o}dinger equation of a 2D hydrogen problem~\cite{Seradjeh2008,Rakhmanov2011,Akzyanov2014,liu2023photonic}, the mass vortex actually supports an infinite number of bound states close to the edges of the mass gap, as plotted in Fig.~\ref{fig:spectrum}(a).
This contrasts with the 1D kink model~(Jackiw-Rebbi), in which the zero mode is the only bound state, due to the tighter confinement of the mass kink compared to the mass vortex.

These bound states are indexed by a principal quantum number $n$ and $(2n+1)$ degenerate angular-momentum numbers $l \in [-n, n]$. The radial spinor components $\boldsymbol{f}(r)$ take the form of generalized Laguerre polynomials multiplied by a power-law factor $r^{|l|}$ and an exponentially decaying tail (see Supplemental Material~[43], Sec.~II). The eigenfrequencies $\omega_n= \pm \frac{2\sqrt{n(n+1)}}{2n+1}m$ and the intensity profiles $|\psi_{n,l}(r)|^2$ for the first few bound states are:
\begin{subequations}
\label{eq:hermitian_solutions}
\begin{align}
%\omega_{n} &= \pm \sqrt{\frac{2n(n+1)}{(2n+1)^2}}m \label{eq:eigenvalues},\\
|\psi_{0}(r)|^2 &= \left|\frac{2 m}{\sqrt{\pi}} (0, 1, -i, 0)^\mathrm{T}e^{-2 m r}\right|^2 = \frac{8m^2}{\pi}e^{-4mr} \label{eq:psi0},\\
|\psi_{1, \pm 1}(r)|^2 &= \frac{2 m^2}{81 \pi}\left(8 m^2 r^2+4 m r+3\right) e^{-\frac{4}{3} m r}.
\label{eq:psi1}
\end{align}
\end{subequations}

As shown in Eq.~\eqref{eq:psi0}, the zero mode has no angular dependence and
resides exclusively in the second and third spinor components: $\boldsymbol{f}(r) = \frac{2m}{\sqrt{\pi}}e^{-2 m r}(0, 1, -i, 0)^{\mathrm{T}}$. For the anti-vortex of $w=-1$, the mode occupies the first and fourth components instead.
Since the zero mode has a much larger spatial decay constant than the rest of the bound states, a practical strategy for achieving single-mode lasing is to introduce a device boundary that absorbs the rest of the modes more than the zero mode.

\textit{Solution of non-Hermitian vortex --- }
We now confine the Hermitian vortex by the absorbing boundary in Eq.~\eqref{eq:boundary}. This non-Hermitian condition constrains the radial spinor components to satisfy $f_1(R) + f_2(R) = f_3(R) + f_4(R) = 0$ at the boundary, and they are Whittaker functions~\cite{S1966} (see Supplemental Material~[43], Sec.~II), a broader class of generalized Laguerre polynomials.
As a result, the boundary induces nonzero components $f_1$, $f_4$ in the zero-mode spinor $\boldsymbol{f}(r)
%= f_2(r)(-i\delta_r, 1, i, -\delta_r)^{\mathrm{T}}
=(f_1, f_2, -if_2, i f_1)^{\mathrm{T}}$.
Here $f_2(r)$ is the usual major component [similar to Eq.~\eqref{eq:psi0}] that decays from the core to the boundary, while $f_1(r)$ is the minor component that matches $-f_2(R)$ at the boundary and decays to zero at the vortex core, as plotted in Supplemental Material~[43], Sec.~III.

The equations are nondimensionalized using the vortex radius $R$ to facilitate scale-invariant analysis, yielding normalized complex eigenvalues $\omega R$ and $\alpha_\parallel R$ that are determined by the normalized real mass $mR$. The complex spectrum of the vortex is plotted in Fig.~\ref{fig:spectrum}(b) for a representative parameter~($mR=1.7$), alongside the Hermitian spectrum in Fig.~\ref{fig:spectrum}(a). Both spectra exhibit the particle-hole symmetry discussed in the above section.
The introduction of the boundary further confines the modes and pushes the frequencies of the nonzero bound states out of the mass gap.
The boundary also lifts the (2$n$ + 1)-fold degeneracy of the angular-momentum states into a series of singlets ($l=0$) and doublets ($\pm l$).

\begin{figure}[!htb]
\centering
\includegraphics[width=0.98\linewidth, keepaspectratio]{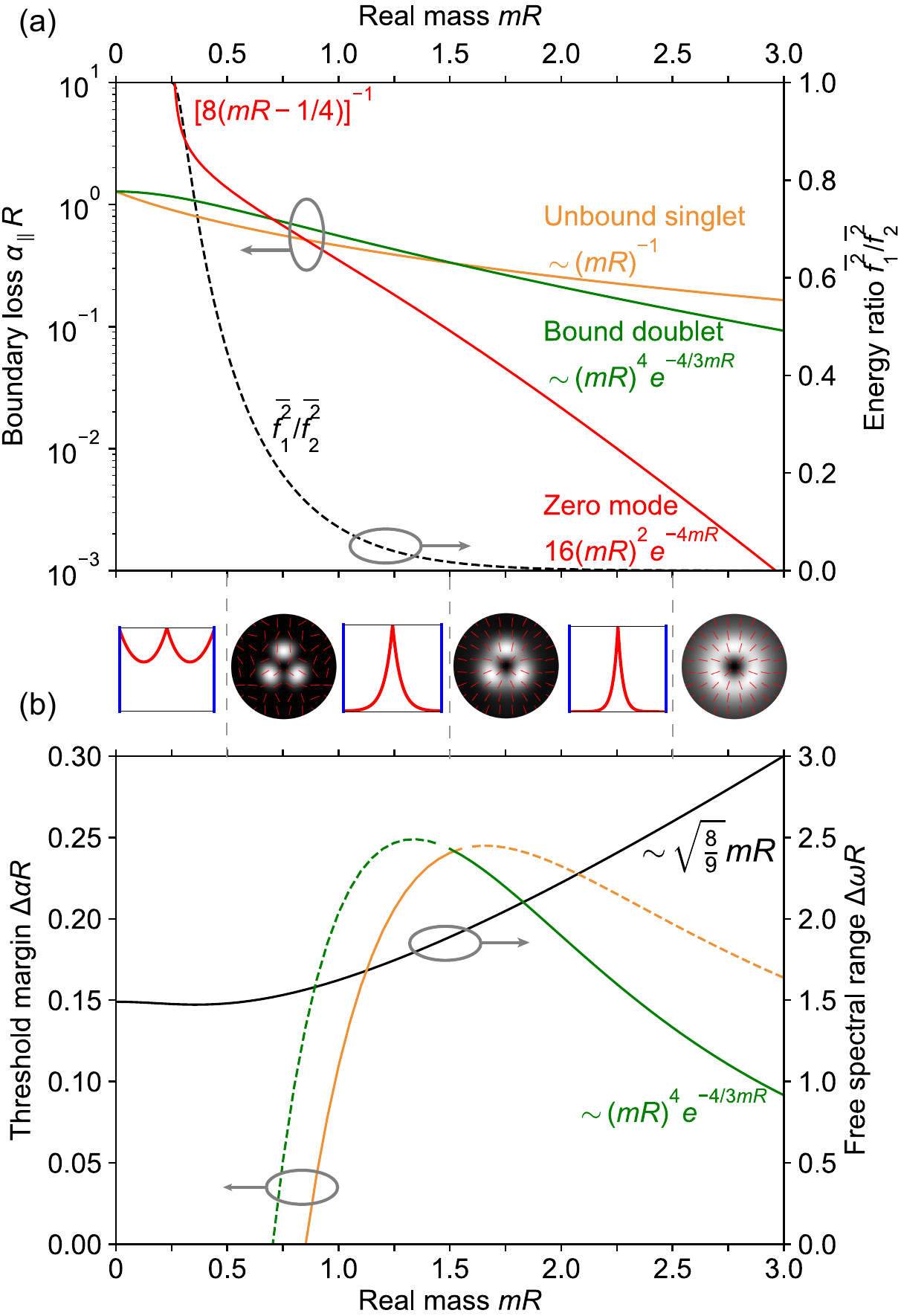}
\caption{Scalings of boundary losses and mode separation. 
(a) Normalized boundary losses $\alpha_\parallel R$ versus the normalized real mass $mR$. 
The zero mode~(red) exhibits an advantageous scaling of $\sim (mR)^2 e^{-4mR}$ for large $mR$, having lower loss than the bound doublet~(green) and unbound singlet~(orange) that are degenerate in the massless limit ($mR \to 0$). 
The black dashed curve~(right axis) plots the minor-to-major component energy ratio $\bar{f_1^2}/\bar{f_2^2}$ of the zero mode. Bottom insets display the evolution of radial intensities and far-field vector-beam patterns at $mR=0.5, 1.5$, and $2.5$, respectively. 
(b) Mode separations in complex eigenvalues. 
The threshold margin $\Delta \alpha R$ is defined by the loss difference between the zero mode and the second lowest-loss mode~(the orange unbound singlet at small $mR$ and the green bound doublet at large $mR$).
The free spectral range $\Delta \omega R$ is plotted as a black curve.}
\label{fig:threshold}
\end{figure}

\begin{figure}[htbp]
\centering
\includegraphics[width=1 \linewidth]{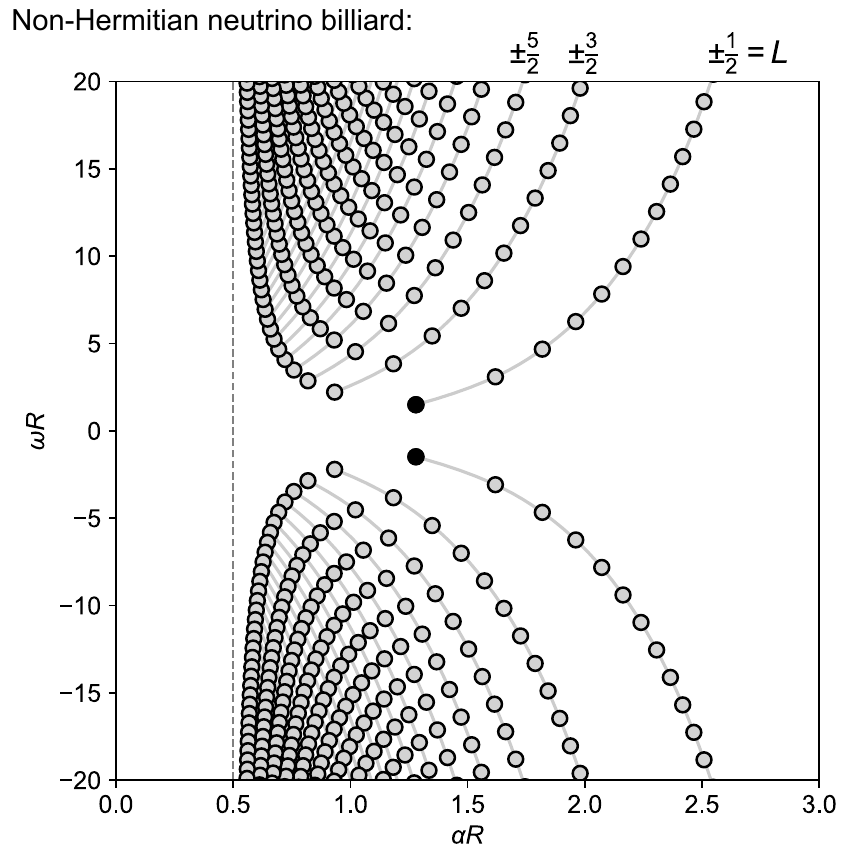}
\caption{Eigenvalue spectrum of the $4\times4$ massless non-Hermitian Dirac vortex.
The complex eigenvalues $E = \omega + i\alpha/2$ are plotted as dimensionless quantities normalized by the vortex radius $R$.
Every data point is 4-fold degenerate for $\pm l$ and both valleys, and the gray lines connect states with the same angular momentum.
The solid black dots highlight the fundamental modes ($l=\pm 1/2$), which could compete with the zero mode for lasing in the massive case.
}
\label{fig:massless_spectrum}
\end{figure}

The boundary loss~($\alpha_\parallel$) is given by the imaginary part of the eigenvalue, which can be estimated from the spatial extent of the wavefunctions in the Hermitian solutions in Eq.~\eqref{eq:hermitian_solutions}: the more localized the mode, the lower the loss.
Since the modal localization decreases with $n$ and increases with $|l|$~(as in the hydrogen problem), the zero mode has the lowest boundary loss~(in the large-$mR$ limit), followed by the ``bound doublet''~($n=1,l=\pm1$). 
Interestingly, the next low-loss singlet, shown as the orange diamond in Fig.~\ref{fig:spectrum}(b), does not correspond to any Hermitian bound state. The origin of this ``unbound singlet'' is revealed in the massless limit of the problem presented in End Matter: it is degenerate with the triplet bound states~($n=1$) and forms the quadruplet of the lowest frequency in Fig.~\ref{fig:massless_spectrum} and in Fig.~\ref{fig:threshold}(a).

\textit{Distinct boundary losses --- }
We plot the normalized losses~$\alpha_\parallel R$ as a function of $mR$ in Fig.~\ref{fig:threshold}(a), and derive their asymptotic scaling behaviors detailed in Supplemental Material~[43], Sec.~II.
The boundary losses of the unbound singlet and bound doublet, degenerate at $mR=0$, bifurcate with distinct scalings for large $mR$.
The loss of the unbound singlet follows a power-law decay $\sim(mR)^{-1}$, because it is not originally bounded in the mass vortex. In contrast, the boundary loss of the bound doublet is exponentially suppressed as $\sim(mR)^4 e^{-\frac{4}{3} m R}$, governed by the decaying tail of the Hermitian wavefunction $|\psi_{1, \pm 1}(r)|^2$ in Eq.~\eqref{eq:psi1}.
%Detailed numerical verifications of these asymptotic scalings are provided in Supplemental Material~[43], Sec.~II.

The loss of the zero mode decays the fastest in the form of $16(mR)^2 e^{-4 m R}$, shown as the solid red line in Fig.~\ref{fig:threshold}(a). In particular, the existence of the finite-sized zero mode requires a threshold of $mR > 1/4$. 
As $mR$ approaches $1/4$, $\alpha_\parallel R=[8(mR - 1/4)]^{-1}$, leading to a divergent boundary loss or a vanishing lifetime for the zero mode.
% For $mR>0.85$, the zero mode has the lowest loss to lase first.
For $mR > 0.85$, the zero mode has the lowest loss to lase first.
Below 0.85, the unbound singlet lases first, as we experimentally confirm in End Matter.
%Detailed analytical derivations are provided in Supplemental Material~[43], Sec.~II.

\textit{Single-mode stability --- }
Stable single-mode operation requires sufficient eigenvalue separation between the lasing mode and the other modes. This separation is quantified in both the real part $\Delta\omega$~(free spectral range) and the imaginary part $\Delta\alpha$~(threshold margin)~\cite{Inoue2025}, as illustrated in Fig.~\ref{fig:spectrum}(b). In this work, $\Delta\alpha = \Delta\alpha_\parallel$, since $\Delta\alpha_\perp$ is identical for all modes as derived in Eq.~\eqref{eq:rad_loss}.

Normalized metrics~($\Delta\omega R$, $\Delta\alpha R$) are plotted as a function of the normalized real mass $mR$ in Fig.~\ref{fig:threshold}(b).
The free spectral range~(black line) is consistently defined by the bound doublet, which increases with $mR$ and approaches $\sqrt{8/9}mR$ from above --- the frequency in the Hermitian limit.
The threshold margin $\Delta \alpha R$, however, is defined by the unbound singlet~(orange line) for small $mR$ and by the bound doublet~(green line) for large $mR$. 
Importantly, $\Delta \alpha R$ 
reaches the global maximum at $mR \approx 1.5$, the crossover of the orange and green lines where the second-lowest-loss mode changes.
% It is advisable to operate the TCSEL with $mR\ge1.5$ to achieve a larger free spectral range and a lower boundary loss.
A practical operating window is around $mR \approx 1.5$ or slightly above, where the threshold margin is near maximal, while the free spectral range and boundary loss are favorable.

\begin{figure}
\centering
\includegraphics[width=0.9\linewidth, keepaspectratio]{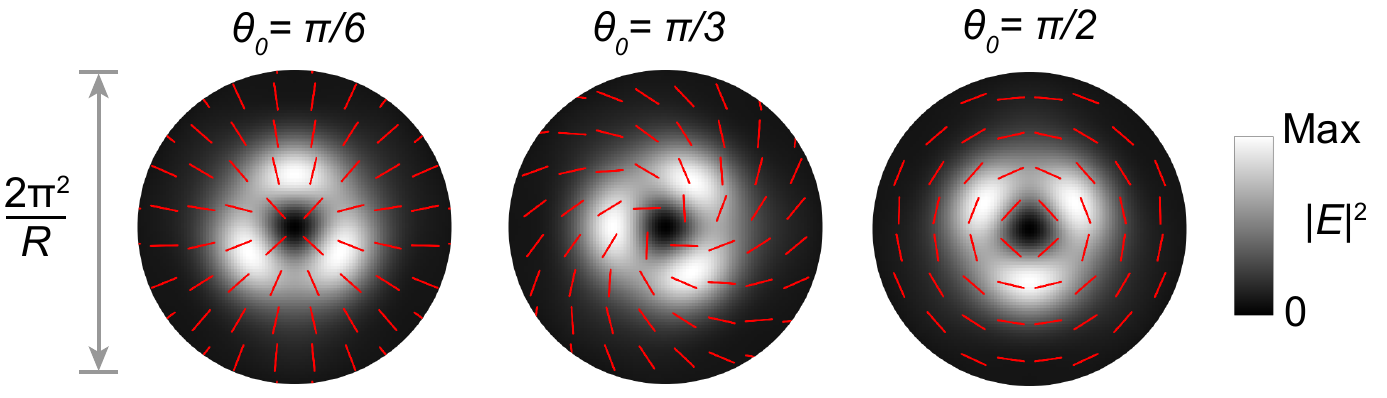}
\caption{Polarization control of the vector-beam output by the initial mass phases $\theta_0$~(at $mR=1.7$). 
%The three-lobe background intensity has its maximum aligned along $-\theta_0$. 
The experimental verifications are presented in Fig.~\ref{fig:experiment_combined} of End Matter.
% The background represents intensity, while red lines indicate the local linear polarization direction at $mR=1.7$.
}
\label{fig:far}
\end{figure}

\textit{Radiation pattern--- }
The far-field pattern is one of the most accessible observables of a laser, especially for identifying the lasing mode.
To obtain the far fields, we derive the radiation operator that maps the spinor wavefunctions of the vortex modes to their corresponding radiative near fields; the far-field patterns in Fig.~\ref{fig:spectrum}(d), Fig.~\ref{fig:threshold} and Fig.~\ref{fig:far} are then the Fourier transforms of these near fields.

Since the spinor wavefunctions are written in the basis of K and K$'$ valley states~(the $C_3$ eigen-states), we express the radiative near field in the basis of the circularly polarized light --- also the eigenstates of the $C_3$ symmetry.
The shared symmetry dictates that the spinor components (1, 3) and (2, 4) couple to the left- and right-circularly polarized (LCP and RCP) light, respectively; the coupling amplitude and a valley-dependent phase are $\sqrt{\mu}$ and $e^{-i\tau_z(\theta+\theta_0)}$, as illustrated in Fig.~\ref{fig:coupling}(a).
This leads to the radiation operator, the $2 \times 4$ matrix in Eq.~\eqref{eq:E_field}, which can be equivalently obtained from the 3D CWT in Supplemental Material~[43], Sec.~III.
\begin{align}
\label{eq:psi}
|\psi_{0}(r,\theta)\rangle &= e^{-i\sigma_z\frac{\theta}{2}}e^{-i\tau_z \frac{(\theta+\theta_0)}{2}}(f_1, f_2, -if_2,  if_1)^{\mathrm{T}}, \\
\label{eq:E_field}
\boldsymbol{E}(r,\theta) &\propto \left(\begin{array}{cccc}
1 & 0 & -1 & 0 \\
0 & -1 & 0 & 1
\end{array}\right)\sqrt{\mu}e^{-i\tau_z(\theta+\theta_0)}|\psi_{0}(r,\theta)\rangle \nonumber \\
&= \sqrt{\mu}f_2(r) \left(\begin{array}{c} ie^{i(\theta+\frac{3}{2}\theta_0)} \\ - e^{-i(\theta+\frac{3}{2}\theta_0)} \end{array}\right) +  \sqrt{\mu}f_1(r) \left(\begin{array}{c}  e^{-i(2\theta+\frac{3}{2}\theta_0)} \\ ie^{i(2\theta+\frac{3}{2}\theta_0)} \end{array}\right).
\end{align}

The two terms in Eq.~\eqref{eq:E_field} reveal that the radiative field of the zero mode is a superposition of two cylindrical vector beams~\cite{Zhan2009}, which are equal-amplitude superpositions of LCP and RCP states with conjugate phases.
The direction of the local linear polarization is determined by half the phase difference between the LCP and RCP states. As a result, the polarization windings of the two vector beams are +1 and -2, which are the two smallest singlet topological charges allowed by the $C_3$ symmetry~(see Supplemental Material~[43], Fig.~S6), the next being +4~\cite{Zhen2014}. Thus, these vector-beam charges are enforced by symmetry and remain invariant for the anti-vortex of $w=-1$, which is merely the mirror image of $w=1$ with reversed winding.

The characteristic three-lobe intensity pattern of TCSEL results from the superposition of the two vector beams with charges +1 and -2.
Such a $C_3$-compatible far field has been experimentally observed in Refs.~\cite{Li2026, yang2022topological,Liu2024,Zhong2026, han2023photonic} and is shown in Fig.~\ref{fig:spectrum}(d), Fig.~\ref{fig:threshold} and Fig.~\ref{fig:far}.
The +1 beam is dominant because the -2 beam is radiated from the minor spinor components~($f_1$) induced by the boundary.
The power ratio of the -2 beam is quantified by $\bar{f_1^2}/\bar{f_2^2} = \int_0^R |f_1|^2 r dr \Big/ \int_0^R |f_2|^2 r dr$, which diminishes rapidly with $mR$ as plotted by the black curve in Fig.~\ref{fig:threshold}(a). At $mR=1.5$, the +1 beam contributes 98\%~($\bar{f_1^2}/\bar{f_2^2}=0.02$).
Therefore, the three lobes evolve into a circular ring with increasing $mR$, as shown in the middle row of Fig.~\ref{fig:threshold}.

\begin{figure*}[t!]
\centering
\includegraphics[width=\linewidth]{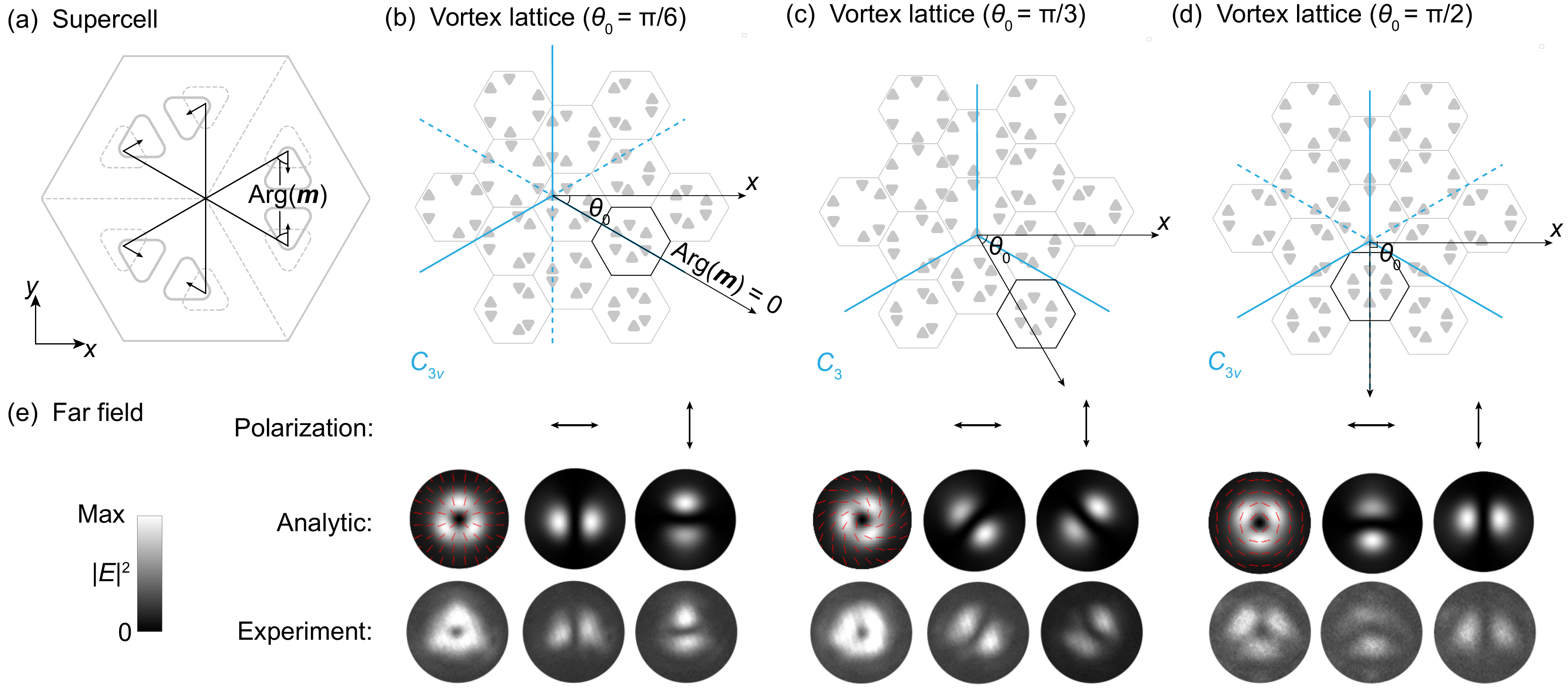}
\caption{Lattice design and experimental validation.
(a) Generalized Kekul\'{e} modulation within the $C_{3v}$ supercell. Dashed and solid gray triangles denote the unperturbed and displaced air holes, respectively. 
(b)--(d) Vortex lattice configurations for initial mass phases $\theta_0 = \pi/6$, $\pi/3$, and $\pi/2$. The initial phase $\theta_0$ is geometrically defined by the angle between the $x$-axis and the spatial direction where $\mathrm{Arg}(\boldsymbol{m})=0$, indicated by black arrows. Blue lines denote the global spatial symmetries of the vortex lattices, manifesting $C_{3v}$ symmetry in (b, d) and reducing to $C_3$ symmetry in (c). 
(e) Far-field vector beams. Theoretical~(top) and experimental~(bottom) total intensities~(left) and intensities filtered by linear polarizers along the indicated axes~($\leftrightarrow$, $\updownarrow$).}
\label{fig:experiment_combined}
\end{figure*}

The polarization of the output vector beam is continuously tunable by varying $\theta_0$, the initial phase of the Dirac mass.
As seen in Eq.~\eqref{eq:E_field}, $\theta_0$ acts as the phase constant for a vector beam, switching the polarization pattern between radial, azimuthal or intermediate spiral patterns shown in Fig.~\ref{fig:far}.
When neglecting the -2 beam, the local polarization direction is $(3\theta_0/2-\pi/4)$ with respect to the local radial direction. 
Accordingly, the three-lobe intensity pattern also rotates with $\theta_0$, with the maximum intensity along $-\theta_0$. 
As a unique capability of the TCSEL, we demonstrate the polarization control experimentally in Fig.~\ref{fig:experiment_combined} of End Matter.

%%%%%%%%%%%%%%%%%%%%%%%%%%% Section_5 %%%%%%%%%%%%%%%%%e%%%%%%%%%%
\textit{Conclusion ---}
We establish a minimal theory for TCSEL, by analytically solving the real-mass~($m$) vortex confined by an infinite well of imaginary potential~(of radius $R$) and perturbatively deriving its radiative properties due to the imaginary mass~($\mu$).
Its bulk Hamiltonian is the non-Hermitian extension of the Jackiw-Rossi model, while the boundary condition is the non-Hermitian counterpart of that in the neutrino-billiard and MIT-bag models.
The predictions are corroborated by experimentally observing the tunable vector-beam polarizations of the zero mode and the unbound singlet.
This non-Hermitian Dirac vortex model sets the theoretical foundation for understanding and developing TCSEL into next-generation diode lasers.

\textit{Acknowledgments --- }
We thank Xicheng Fan, Tianwei Zheng, Boyuan Liu, Xiaoqi Sun and Shu Chen for discussions. 
This work was supported by the CAS through the Project for Young Scientists in Basic Research (Grant No.~YSBR-021); 
the National Natural Science Foundation of China (Grant Nos.~12574425 to L.Y. and 12125405 to Z.W.); 
the National Key R\&D Program of China (Grant No.~2023YFA1406702 to Z.W.); 
and the IOP-HKUST Joint Laboratory for Wave Functional Materials Research. 
This work was also supported by the Synergetic Extreme Condition User Facility and the Laboratory of Microfabrication, IOP CAS.

\nocite{SupplementalMaterialFullManuscript}
\bibliography{references}

@article{Li2026,
   author = {Haitao Li and Jiusi Yu and Jiayu Fan and Shijie Kang and Bo Hou and Zhi‐Kang Lin and Hanchuan Chen and Jian‐Hua Jiang and Xiaoxiao Wu},
   doi = {10.1002/lpor.202502844},
   issn = {1863-8880},
   journal = {Laser \& Photonics Reviews},
   month = {1},
   title = {Unveiling Spin‐Orbital Angular Momentum Locking in Photonic Dirac Vortex Cavities},
   year = {2026}
}

@article{S1966,
   author = {D. S. and Milton Abramowitz and Irene A. Stegun},
   doi = {10.2307/2004284},
   issn = {00255718},
   issue = {93},
   journal = {Mathematics of Computation},
   month = {1},
   pages = {167},
   title = {Handbook of Mathematical Functions with Formulas, Graphs, and Mathematical Tables.},
   volume = {20},
   year = {1966}
}

@article{Engquist1977,
   author = {Bjorn Engquist and Andrew Majda},
   doi = {10.1090/S0025-5718-1977-0436612-4},
   issn = {0025-5718},
   issue = {139},
   journal = {Mathematics of Computation},
   pages = {629-651},
   title = {Absorbing boundary conditions for the numerical simulation of waves},
   volume = {31},
   year = {1977}
}

@article{Inoue2025,
   author = {Takuya Inoue and Kentaro Maeda and Masahiro Yoshida and John Gelleta and Shumpei Katsuno and Kenji Ishizaki and Menaka De Zoysa and Susumu Noda},
   doi = {10.1109/JSTQE.2024.3452126},
   issn = {1077-260X},
   issue = {2: Pwr. and Effic. Scaling in},
   journal = {IEEE Journal of Selected Topics in Quantum Electronics},
   month = {3},
   pages = {1-8},
   title = {Influence of Band-Edge Frequency Non-Uniformity in Ultra-Large-Area Photonic-Crystal Surface-Emitting Lasers},
   volume = {31},
   year = {2025}
}

@article{Zhong2026,
   author = {Mou Zhong and Xiaoqiong Bi and Mengyuan Song and Nanli Mou and Delin Zhang and Xiaolu Zhuo and Jingtian Hu and Biye Xie and Xianyu Ao and Jun Guan},
   doi = {10.1038/s41467-026-70833-1},
   issn = {2041-1723},
   journal = {Nature Communications},
   month = {3},
   title = {Plasmonic Dirac-vortex lasers via three-dimensional photonic mass vortices engineering},
   year = {2026}
}

@article{Liu2024,
   author = {Junhong Liu and Yunfei Xu and Rusong Li and Yongqiang Sun and Kaiyao Xin and Jinchuan Zhang and Quanyong Lu and Ning Zhuo and Junqi Liu and Lijun Wang and Fengmin Cheng and Shuman Liu and Fengqi Liu and Shenqiang Zhai},
   doi = {10.1038/s41467-024-48788-y},
   issn = {2041-1723},
   issue = {1},
   journal = {Nature Communications},
   month = {6},
   pages = {4431},
   title = {High-power electrically pumped terahertz topological laser based on a surface metallic Dirac-vortex cavity},
   volume = {15},
   year = {2024}
}

@book{Coldren2012,
   author = {Larry A. Coldren and Scott W. Corzine and Milan L. Ma{\v{s}}anovi{\'c}},
   doi = {10.1002/9781118148167},
   isbn = {9780470484128},
   month = {3},
   publisher = {Wiley},
   title = {Diode Lasers and Photonic Integrated Circuits},
   year = {2012}
}

@article{Noda2024,
   author = {Susumu Noda and Masahiro Yoshida and Takuya Inoue and Menaka De Zoysa and Kenji Ishizaki and Ryoichi Sakata},
   doi = {10.1038/s44287-024-00113-x},
   issn = {2948-1201},
   issue = {12},
   journal = {Nature Reviews Electrical Engineering},
   month = {11},
   pages = {802-814},
   title = {Photonic-crystal surface-emitting lasers},
   volume = {1},
   year = {2024}
}

@article{Zhen2014,
   author = {Bo Zhen and Chia Wei Hsu and Ling Lu and A. Douglas Stone and Marin Solja{\v{c}}i{\'c}},
   doi = {10.1103/PhysRevLett.113.257401},
   issn = {0031-9007},
   issue = {25},
   journal = {Physical Review Letters},
   month = {12},
   pages = {257401},
   title = {Topological Nature of Optical Bound States in the Continuum},
   volume = {113},
   year = {2014}
}

@article{Zhan2009,
   author = {Qiwen Zhan},
   doi = {10.1364/AOP.1.000001},
   issn = {1943-8206},
   issue = {1},
   journal = {Advances in Optics and Photonics},
   month = {1},
   pages = {1},
   title = {Cylindrical vector beams: from mathematical concepts to applications},
   volume = {1},
   year = {2009}
}

@article{Seradjeh2008,
   author = {B. Seradjeh},
   doi = {10.1016/j.nuclphysb.2008.07.011},
   issn = {05503213},
   issue = {1-2},
   journal = {Nuclear Physics B},
   month = {12},
   pages = {182-189},
   title = {Midgap spectrum of the fermion–vortex system},
   volume = {805},
   year = {2008}
}

@article{Ge2013,
   author = {Li Ge and Hakan E. Türeci},
   doi = {10.1103/PhysRevA.88.053810},
   issn = {1050-2947},
   issue = {5},
   journal = {Physical Review A},
   month = {11},
   pages = {053810},
   title = {Antisymmetric  PT  -photonic structures with balanced positive- and negative-index materials},
   volume = {88},
   year = {2013}
}

@article{Rakhmanov2011,
   author = {A. L. Rakhmanov and A. V. Rozhkov and Franco Nori},
   doi = {10.1103/PhysRevB.84.075141},
   issn = {1098-0121},
   issue = {7},
   journal = {Physical Review B},
   month = {8},
   pages = {075141},
   title = {Majorana fermions in pinned vortices},
   volume = {84},
   year = {2011}
}

@article{Kawabata2019,
   author = {Kohei Kawabata and Sho Higashikawa and Zongping Gong and Yuto Ashida and Masahito Ueda},
   doi = {10.1038/s41467-018-08254-y},
   issn = {2041-1723},
   issue = {1},
   journal = {Nature Communications},
   month = {1},
   pages = {297},
   title = {Topological unification of time-reversal and particle-hole symmetries in non-Hermitian physics},
   volume = {10},
   year = {2019}
}

@article{Tumulka2016,
   author = {Roderich Tumulka},
   month = {1},
   title = {Detection Time Distribution for Dirac Particles},
   year = {2016},
   journal = {arXiv.1601.04571},
   url = {https://arxiv.org/abs/1601.04571}
}

@article{Kazarinov1985,
   author = {R. Kazarinov and C. Henry},
   doi = {10.1109/JQE.1985.1072627},
   issn = {0018-9197},
   issue = {2},
   journal = {IEEE Journal of Quantum Electronics},
   month = {2},
   pages = {144-150},
   title = {Second-order distributed feedback lasers with mode selection provided by first-order radiation losses},
   volume = {21},
   year = {1985}
}

@article{Kogelnik1972,
   author = {H. Kogelnik and C. V. Shank},
   doi = {10.1063/1.1661499},
   issn = {0021-8979},
   issue = {5},
   journal = {Journal of Applied Physics},
   month = {5},
   pages = {2327-2335},
   title = {Coupled-Wave Theory of Distributed Feedback Lasers},
   volume = {43},
   year = {1972}
}

@article{Liang2012,
   author = {Yong Liang and Chao Peng and Kyosuke Sakai and Seita Iwahashi and Susumu Noda},
   doi = {10.1364/OE.20.015945},
   issn = {1094-4087},
   issue = {14},
   journal = {Optics Express},
   month = {7},
   pages = {15945},
   title = {Three-dimensional coupled-wave analysis for square-lattice photonic crystal surface emitting lasers with transverse-electric polarization: finite-size effects},
   volume = {20},
   year = {2012}
}

@article{Fu2008,
   author = {Liang Fu and C. L. Kane},
   doi = {10.1103/PhysRevLett.100.096407},
   issn = {0031-9007},
   issue = {9},
   journal = {Physical Review Letters},
   month = {3},
   pages = {096407},
   title = {Superconducting Proximity Effect and Majorana Fermions at the Surface of a Topological Insulator},
   volume = {100},
   year = {2008}
}

@article{Teo2010,
   author = {Jeffrey C. Y. Teo and C. L. Kane},
   doi = {10.1103/PhysRevLett.104.046401},
   issn = {0031-9007},
   issue = {4},
   journal = {Physical Review Letters},
   month = {1},
   pages = {046401},
   title = {Majorana Fermions and Non-Abelian Statistics in Three Dimensions},
   volume = {104},
   year = {2010}
}

@article{Akhmerov2008,
   author = {A. R. Akhmerov and C. W. J. Beenakker},
   doi = {10.1103/PhysRevB.77.085423},
   issn = {1098-0121},
   issue = {8},
   journal = {Physical Review B},
   month = {2},
   pages = {085423},
   title = {Boundary conditions for Dirac fermions on a terminated honeycomb lattice},
   volume = {77},
   year = {2008}
}

@article{Antoine2017,
   author = {X. Antoine and E. Lorin and Q. Tang},
   doi = {10.1080/00268976.2017.1290834},
   issn = {0026-8976},
   issue = {15-16},
   journal = {Molecular Physics},
   month = {8},
   pages = {1861-1879},
   title = {A friendly review of absorbing boundary conditions and perfectly matched layers for classical and relativistic quantum waves equations},
   volume = {115},
   year = {2017}
}

@article{Katsnelson2006,
   author = {M. I. Katsnelson and K. S. Novoselov and A. K. Geim},
   doi = {10.1038/nphys384},
   issn = {1745-2473},
   issue = {9},
   journal = {Nature Physics},
   month = {9},
   pages = {620-625},
   title = {Chiral tunnelling and the Klein paradox in graphene},
   volume = {2},
   year = {2006}
}

@article{Hou2007,
   author = {Chang-Yu Hou and Claudio Chamon and Christopher Mudry},
   doi = {10.1103/PhysRevLett.98.186809},
   issn = {0031-9007},
   issue = {18},
   journal = {Physical Review Letters},
   month = {5},
   pages = {186809},
   title = {Electron Fractionalization in Two-Dimensional Graphenelike Structures},
   volume = {98},
   year = {2007}
}

@article{Chodos1974,
   author = {A. Chodos and R. L. Jaffe and K. Johnson and C. B. Thorn and V. F. Weisskopf},
   doi = {10.1103/PhysRevD.9.3471},
   issn = {0556-2821},
   issue = {12},
   journal = {Physical Review D},
   month = {6},
   pages = {3471-3495},
   title = {New extended model of hadrons},
   volume = {9},
   year = {1974}
}

@book{morthier2013handbook,
  title={Handbook of distributed feedback laser diodes},
  author={Morthier, Geert and Vankwikelberge, Patrick},
  year={2013},
  publisher={Artech House}
}

@article{Christensen2014,
   author = {Thomas Christensen and Weihua Wang and Antti-Pekka Jauho and Martijn Wubs and N. Asger Mortensen},
   doi = {10.1103/PhysRevB.90.241414},
   issn = {1098-0121},
   issue = {24},
   journal = {Physical Review B},
   month = {12},
   pages = {241414},
   title = {Classical and quantum plasmonics in graphene nanodisks: Role of edge states},
   volume = {90},
   year = {2014}
}

@article{Jackiw1981,
   author = {R. Jackiw and P. Rossi},
   doi = {10.1016/0550-3213(81)90044-4},
   issn = {05503213},
   issue = {4},
   journal = {Nuclear Physics B},
   month = {12},
   pages = {681-691},
   title = {Zero modes of the vortex-fermion system},
   volume = {190},
   year = {1981}
}

@article{Berry1987neutrino,
   author = {Michael Victor Berry and R. J. Mondragon},
   doi = {10.1098/rspa.1987.0080},
   issn = {0080-4630},
   issue = {1842},
   journal = {Proceedings of the Royal Society of London. A. Mathematical and Physical Sciences},
   month = {7},
   pages = {53-74},
   title = {Neutrino billiards: time-reversal symmetry-breaking without magnetic fields},
   volume = {412},
   year = {1987}
}

@article{Jackiw1976,
   author = {R. Jackiw and C. Rebbi},
   doi = {10.1103/PhysRevD.13.3398},
   issn = {0556-2821},
   issue = {12},
   journal = {Physical Review D},
   month = {6},
   pages = {3398-3409},
   title = {Solitons with fermion number ½},
   volume = {13},
   year = {1976}
}

@article{Cheng2024,
   author = {Hengbin Cheng and Jingyu Yang and Zhong Wang and Ling Lu},
   doi = {10.1038/s41467-024-51670-6},
   issn = {2041-1723},
   issue = {1},
   journal = {Nature Communications},
   month = {8},
   pages = {7327},
   title = {Observation of monopole topological mode},
   volume = {15},
   year = {2024}
}

@article{Akzyanov2014,
   author = {R. S. Akzyanov and A. V. Rozhkov and A. L. Rakhmanov and Franco Nori},
   doi = {10.1103/PhysRevB.89.085409},
   issn = {1098-0121},
   issue = {8},
   journal = {Physical Review B},
   month = {2},
   pages = {085409},
   title = {Tunneling spectrum of a pinned vortex with a robust Majorana state},
   volume = {89},
   year = {2014}
}

@article{Gao2020dirac,
   author = {Xiaomei Gao and Lechen Yang and Hao Lin and Lang Zhang and Jiafang Li and Fang Bo and Zhong Wang and Ling Lu},
   doi = {10.1038/s41565-020-0773-7},
   issn = {1748-3387},
   issue = {12},
   journal = {Nature Nanotechnology},
   month = {12},
   pages = {1012-1018},
   title = {Dirac-vortex topological cavities},
   volume = {15},
   year = {2020}
}

@article{Yang2022topological,
   author = {Lechen Yang and Guangrui Li and Xiaomei Gao and Ling Lu},
   doi = {10.1038/s41566-022-00972-6},
   issn = {1749-4885},
   issue = {4},
   journal = {Nature Photonics},
   month = {4},
   pages = {279-283},
   title = {Topological-cavity surface-emitting laser},
   volume = {16},
   year = {2022}
}

@article{Ma2023room,
   author = {Jingwen Ma and Taojie Zhou and Mingchu Tang and Haochuan Li and Zhan Zhang and Xiang Xi and Mickael Martin and Thierry Baron and Huiyun Liu and Zhaoyu Zhang and Siming Chen and Xiankai Sun},
   doi = {10.1038/s41377-023-01290-4},
   issn = {2047-7538},
   issue = {1},
   journal = {Light: Science \& Applications},
   month = {10},
   pages = {255},
   title = {Room-temperature continuous-wave topological Dirac-vortex microcavity lasers on silicon},
   volume = {12},
   year = {2023}
}

@article{Han2023photonic,
   author = {Song Han and Yunda Chua and Yongquan Zeng and Bofeng Zhu and Chongwu Wang and Bo Qiang and Yuhao Jin and Qian Wang and Lianhe Li and Alexander Giles Davies and Edmund Harold Linfield and Yidong Chong and Baile Zhang and Qi Jie Wang},
   doi = {10.1038/s41467-023-36418-y},
   issn = {2041-1723},
   issue = {1},
   journal = {Nature Communications},
   month = {2},
   pages = {707},
   title = {Photonic Majorana quantum cascade laser with polarization-winding emission},
   volume = {14},
   year = {2023}
}

@book{padullaparthi2021vcsel,
  title={VCSEL industry: communication and sensing},
  author={Padullaparthi, Babu Dayal and Tatum, Jim and Iga, Kenichi},
  year={2021},
  publisher={John Wiley \& Sons}
}

@article{Inoue2022general,
   author = {Takuya Inoue and Masahiro Yoshida and John Gelleta and Koki Izumi and Keisuke Yoshida and Kenji Ishizaki and Menaka De Zoysa and Susumu Noda},
   doi = {10.1038/s41467-022-30910-7},
   issn = {2041-1723},
   issue = {1},
   journal = {Nature Communications},
   month = {7},
   pages = {3262},
   title = {General recipe to realize photonic-crystal surface-emitting lasers with 100-W-to-1-kW single-mode operation},
   volume = {13},
   year = {2022}
}

@article{Yoshida2023high,
   author = {Masahiro Yoshida and Shumpei Katsuno and Takuya Inoue and John Gelleta and Koki Izumi and Menaka De Zoysa and Kenji Ishizaki and Susumu Noda},
   doi = {10.1038/s41586-023-06059-8},
   issn = {0028-0836},
   issue = {7966},
   journal = {Nature},
   month = {6},
   pages = {727-732},
   title = {High-brightness scalable continuous-wave single-mode photonic-crystal laser},
   volume = {618},
   year = {2023}
}

@article{Gao2025far,
   author = {Xiaomei Gao and Siqi Chang and Zongliang Li and Tianrui Zhai},
   doi = {10.1364/OL.539008},
   issn = {0146-9592},
   issue = {2},
   journal = {Optics Letters},
   month = {1},
   pages = {245},
   title = {Far fields of two-dimensional TM modes in Dirac-vortex topological cavity},
   volume = {50},
   year = {2025}
}

@article{Liang2013three,
   author = {Yong Liang and Chao Peng and Kenji Ishizaki and Seita Iwahashi and Kyosuke Sakai and Yoshinori Tanaka and Kyoko Kitamura and Susumu Noda},
   doi = {10.1364/OE.21.000565},
   issn = {1094-4087},
   issue = {1},
   journal = {Optics Express},
   month = {1},
   pages = {565},
   title = {Three-dimensional coupled-wave analysis for triangular-lattice photonic-crystal surface-emitting lasers with transverse-electric polarization},
   volume = {21},
   year = {2013}
}

@book{liu2023photonic,
  title={Photonic Crystal Cavity Systems Described by Dirac Equation},
  author={Liu, Boyuan},
  year={2023},
  publisher={The University of Wisconsin-Madison}
}

@misc{SupplementalMaterialFullManuscript,
  title={{Supplemental Material in the full manuscript}},
  note={{See the full manuscript version for Supplemental Material.}},
  year={2026}
}

\begin{comment}

\end{comment}

\clearpage

\begin{center}
\textbf{END MATTER}
\end{center}

\begin{center}
\textbf{Non-Hermitian ``Neutrino Billiard''}
\end{center}

%\textit{Non-Hermitian ``Neutrino billiard'' --- } 
We solve the non-Hermitian Dirac vortex in the massless limit~($m=\mu=0$), namely a Dirac fermion confined by the infinite imaginary potential of radius $R$.
This differs from the ``neutrino billiard'' model~\cite{berry1987neutrino},
which employs a Hermitian hard wall of infinite real mass, as summarized in Table~\ref{tab:scalings}.

\textit{Characteristic equation --- }
In the massless limit, the $4\times4$ Hamiltonian decouples into two independent $2\times2$ Dirac cones at the $\mathrm{K}$ and $\mathrm{K}'$ valleys.
For a single valley, the generalized angular momentum operator is $L_z = -i\partial_{\theta} + \frac{1}{2}\sigma_z$~($w=0$), whose eigenvalue $l$ takes half-integer values ($l=\pm 1/2, \pm 3/2, \dots$).
The coupled radial equations (see Supplemental Material~[43], Sec.~II.C) can be analytically decoupled into the standard Bessel differential equations for the complex energy $E = \omega + i\alpha/2$. 
The radial spinor is thus given by:
\begin{equation}
f_1(r) = A J_{l-1/2}(\mathcal{Z}\frac{r}{R}), \quad f_2(r) = i A J_{l+1/2}(\mathcal{Z}\frac{r}{R}),
\end{equation}
where $\mathcal{Z} = 2E R$ is the dimensionless complex frequency. Imposing the non-Hermitian boundary condition $f_1+f_2=0$ at $r=R$ yields the characteristic equation:
\begin{equation}
J_{l-1/2}(\mathcal{Z}) + i J_{l+1/2}(\mathcal{Z}) = 0. \label{eq:root_K}
\end{equation}

\begin{table}[b]
\caption{\label{tab:scalings} The massless limit of the non-Hermitian Dirac vortex problem can be viewed as the non-Hermitian version of the ``neutrino billiard'' model proposed by Berry and Mondragon~\cite{berry1987neutrino}. $|\psi\rangle$ denotes the two-component Dirac spinor of a single valley.}
\begin{ruledtabular}
\begin{tabular}{c|c|c}
Neutrino billiard     & Hermitian & Non-Hermitian \\ \hline
Boundary & Infinite mass & Infinite imaginary potential \\
condition& (hard-wall boundary) & (absorbing boundary)\\
 & $-{i}\sigma_z(\boldsymbol{\sigma} \cdot \hat{n}) |\psi\rangle=|\psi\rangle$ & $-(\boldsymbol{\sigma} \cdot \hat{n}) |\psi\rangle=|\psi\rangle$\\\hline
Eigenvalue & \multirow{2}{*}{$J_{l-\frac{1}{2}}(\mathcal{Z}) =  J_{l+\frac{1}{2}}(\mathcal{Z}) $}& \multirow{2}{*}{$J_{l-\frac{1}{2}}(\mathcal{Z}) = -{i}J_{l+\frac{1}{2}}(\mathcal{Z})$} \\
equation & & \\
%&  Berry-Mondragon 1987& Massless limit of this work\\
\end{tabular}
\end{ruledtabular}
\end{table}

\textit{Eigenvalue degeneracy --- }
Equation~\eqref{eq:root_K} is invariant under $l \to -l$ due to the Bessel parity relation $J_{-n}(z) = (-1)^n J_n(z)$, rendering all complex eigenvalues doubly degenerate~(fourfold when including both valleys), as plotted in Fig.~\ref{fig:massless_spectrum}. The fundamental mode ($l=\pm1/2$, solid black dots) corresponds to $J_0(\mathcal{Z}) + i J_1(\mathcal{Z}) = 0$. This fourfold state separates into the unbound singlet and bound triplet in the main text.

\textit{Lower bound on loss --- }
Figure~\ref{fig:massless_spectrum} reveals that the boundary loss $\alpha R$ decreases with $|l|$ and increases with $|\omega|$. In the large-$|l|$ limit, the modes form whispering-gallery orbits. The vanishing single-bounce boundary transmittance balances the divergent collision frequency, yielding a nonzero asymptotic loss bound as shown in Fig.~\ref{fig:massless_spectrum}. Expanding Eq.~\eqref{eq:root_K} for $l \to \infty$ (see Supplemental Material~[43], Sec.~II.H) confirms that 
\begin{equation}
\lim_{l \to \infty} \alpha R = \lim_{l \to \infty} \mathrm{Im}(\mathcal{Z}) = 0.5.
\end{equation}
This lower bound on the boundary loss suggests that the zero mode could have the lowest loss in the mass vortex when its boundary loss $\alpha_\parallel R<0.5$.

\begin{center}
\textbf{Experimental Verifications}
\end{center}

To experimentally validate our theoretical predictions, we implement the non-Hermitian Dirac vortex in optically pumped topological-cavity surface-emitting lasers, following our previous work~\cite{yang2022topological}.

\textit{$C_{3v}$-supercell design --- }
We introduce a new refinement to the generalized Kekul\'{e} modulation of the honeycomb-lattice supercell shown in Fig.~\ref{fig:experiment_combined}(a).
In Refs.~\cite{gao2020dirac,yang2022topological}, we displaced only one sublattice of the three triangular holes, and the modulated supercell always has $C_3$ symmetry for 2$\pi$ modulation phases.
Here, we symmetrically displace both sublattices in the supercell that preserves the $C_{3v}$ symmetry for all modulation angles.
The additional mirror symmetry eliminates the chiral mass~($m_3$), representing the asymmetry between the two sublattices, so that $m_3 = 0$.
Since $m_3$ is present in the primitive unit cell which is inherently nonradiative, there is no corresponding $\mu_3$ term in the first place.

\textit{Initial mass phase $\theta_0$ --- }
We define $\mathrm{Arg}(\boldsymbol{m})=0$ when the six air holes move towards the center of the supercell shown in Fig.~\ref{fig:experiment_combined}(a).
When constructing the vortex, we have the freedom of where to place the $\mathrm{Arg}(\boldsymbol{m})=0$ mass.
This angular freedom is the initial mass phase $\theta_0$. 
Figures~\ref{fig:experiment_combined}(b)--(d) illustrate three vortex lattices for $\theta_0 = \pi/6, \pi/3,$ and $\pi/2$, respectively. Notably, the vortex symmetry is determined by $\theta_0$: the vortices with $\theta_0 = \pi/6$ and $\pi/2$ preserve 
$C_{3v}$ symmetry while the intermediate $\theta_0$ values reduce the spatial symmetry to $C_3$.

\textit{Far-field polarization --- } 
In Fig.~\ref{fig:experiment_combined}(e), we compare the experimental and the theoretical far-field patterns for three initial mass phases ($\theta_0 = \pi/6, \pi/3$, and $\pi/2$).
The total unpolarized three-lobe intensity has its maxima aligned along the $-\theta_0$ direction~($\mathrm{Arg}(\boldsymbol{m})=0$).
When filtered through a horizontal or vertical linear polarizer, the far fields display polarization directions rotating with $\theta_0$ as predicted. 
These results can be understood from symmetry. In $C_{3v}$ vortices of $\theta_0 = \pi/6$ and $\pi/2$, the polarization must respect the mirror symmetry, yielding either radial or azimuthal polarizations. In $C_{3}$ vortices of other $\theta_0$ values in between, the polarization is of the spiral shape.

\begin{figure}[htbp]
\centering
\includegraphics[width=0.95\linewidth]{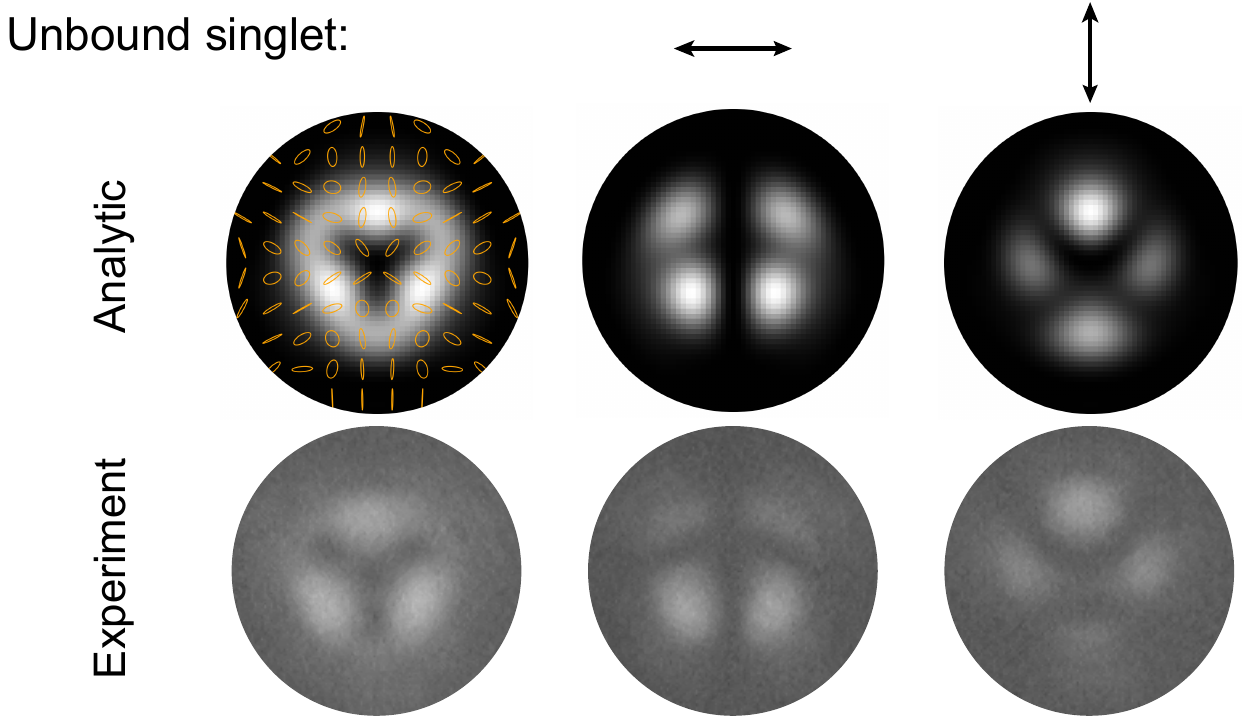}
\caption{Observation of the unbound singlet. The lasing mode switches from the zero mode to the unbound singlet, as we shrink the pump spot~(corresponding to a smaller $mR$). The far fields agrees with the theoretical predictions.}
\label{fig:unbound}
\end{figure}

\textit{Observation of unbound singlet --- }
As predicted in Fig.~\ref{fig:threshold}(a), the boundary loss of the zero mode exceeds that of the unbound singlet for $mR < 0.85$, leaving the latter the lowest-threshold mode~(also see Supplemental Material~[43], Fig.~S2). Here, we experimentally confirm this crossover by reducing the optical pumping area, which effectively decreases $mR$. Taking a device with $\theta_0 = \pi/2$ as an example, this reduction triggers a clear lasing transition. As shown in Fig.~\ref{fig:unbound}, the experimentally observed far field after reducing the pumped region agrees well with the theoretical prediction of the unbound singlet.

Altogether, the experimentally observed tunable vector beam and the crossover between lasing modes verify our minimal theory for TCSEL.

\end{document}